\documentclass[12pt,preprint]{aastex}

\newcommand\lsim{\mathrel{\spose{\lower 3pt\hbox{$\mathchar"218$}}
     \raise 2.0pt\hbox{$\mathchar"13C$}}}
\newcommand\gsim{\mathrel{\spose{\lower 3pt\hbox{$\mathchar"218$}}
     \raise 2.0pt\hbox{$\mathchar"13E$}}}
\def\ltsima{$\; \buildrel < \over \sim \;$}
\def\lsim{\lower.5ex\hbox{\ltsima}}
\def\gtsima{$\; \buildrel > \over \sim \;$}
\def\gsim{\lower.5ex\hbox{\gtsima}}

\def\apjl{ApJL}
\def\apjs{ApJSS}

\def\aap{A\&A}

\def\lat{{\it Fermi}-LAT}

\usepackage{lineno}
\slugcomment{}

\shorttitle{MeV NuSTAR Blazar}
\shortauthors{Ajello et al.}

\usepackage{calrsfs,euscript,mathrsfs,amssymb}

\begin{document}


\title{{\it NuSTAR, Swift,} and GROND observations 
of the flaring MeV blazar: PMN~J0641$-$0320}
%The MeV blazar PMN~J0641$-$0320 observed by NuSTAR}


\email{majello@clemson.edu}

%%%%%%%%%%%%%%%%%%%%%%%%%%%%%%%%%%%%%%%%%%%%%%%%%%%%%%%%%%%%%%%%%%
%
%  history of changes:
%                     May.1st.2014: created
%%                    
%
%
%
%%%%%%%%%%%%%%%%%%%%%%%%%%%%%%%%%%%%%%%%%%%%%%%%%%%%%%%%%%%%%%%%%%%

\author{
M.~Ajello\altaffilmark{1},
G.~Ghisellini\altaffilmark{2}, 
V.~S.~Paliya\altaffilmark{1,3},
D.~Kocevski\altaffilmark{4},
G.~Tagliaferri\altaffilmark{2},
G.~Madejski\altaffilmark{5},
A.~Rau\altaffilmark{6},
P.~Schady\altaffilmark{6},
J.~Greiner\altaffilmark{6},
F.~Massaro\altaffilmark{7},
M.~Balokovi\'{c}\altaffilmark{8},
R.~B\"uhler\altaffilmark{9},
M.~Giomi\altaffilmark{9},
L.~Marcotulli\altaffilmark{1},
F.~D'Ammando\altaffilmark{10,11}
% NuSTAR team
D.\ Stern\altaffilmark{12}, 
S.\ E.\ Boggs\altaffilmark{13}, 
F.\ E.\ Christensen\altaffilmark{14}, 
W.\ W.\ Craig\altaffilmark{14,15}, 
C.\ J.\ Hailey\altaffilmark{16}, 
F.\ A.\ Harrison\altaffilmark{8}, 
W.\ W.\ Zhang\altaffilmark{4} 
}
\altaffiltext{1}{Department of Physics and Astronomy, Clemson University, Kinard Lab of Physics, Clemson, SC 29634-0978, USA}
\altaffiltext{2}{INAF -- Osservatorio Astronomico di Brera, via
  E. Bianchi 46, I--23807 Merate, Italy}
  \altaffiltext{3}{Indian Institute of Astrophysics, Block II Koramangala, Bangalore, India, 560034}
\altaffiltext{4}{NASA Goddard Space Flight Center, Greenbelt, MD
  20771, USA}
\altaffiltext{5}{Kavli Institute for Particle Astrophysics and Cosmology,
			SLAC National Accelerator Laboratory, Menlo Park, CA 94025, USA}

\altaffiltext{6}{Max--Planck--Institut f\"ur extraterrestrische Physik, Giessenbachstrasse 1, 
			85748, Garching, Germany}
\altaffiltext{7}{Dipartimento di Fisica, Universit\`a degli Studi di Torino, via Pietro Giuria 1, I-10125 Torino, Italy}

\altaffiltext{8}{Cahill Center for Astronomy and Astrophysics, California Institute
			of Technology, Pasadena, CA 91125, USA}

\altaffiltext{9}{Deutsches Elektronen Zeuthen, Synchrotron
DESY, D-15738 Zeuthen, Germany}

\altaffiltext{10}{INAF Istituto di Radioastronomia, I-40129 Bologna,
  Italy}
\altaffiltext{11}{Dipartimento di Astronomia, Universit\`a di Bologna, I- 40127 Bologna, Italy}

\altaffiltext{12}{Jet Propulsion Laboratory, California Institute of Technology, Pasadena, 
			CA 91109, USA}
\altaffiltext{13}{Space Sciences Laboratory, University of California,
  Berkeley, CA 94720, USA}

\altaffiltext{14}{DTU Space - National Space Institute, Technical University of
			Denmark, Elektrovej 327, 2800 Lyngby, Denmark}

\altaffiltext{15}{Lawrence Livermore National Laboratory, Livermore, CA 94550, USA}
\altaffiltext{16}{Columbia Astrophysics Laboratory, Columbia University, New York, 
			NY 10027, USA} 

\begin{abstract}
MeV blazars are a sub--population of the blazar family, exhibiting 
larger--than--average jet powers, accretion luminosities and
black hole masses. Because of their extremely hard X--ray continua, these objects
are best studied in the X-ray domain.
Here, we report on the discovery by the {\it Fermi} Large Area Telescope and
subsequent follow-up observations with {\it NuSTAR}, {\it Swift} and GROND
of a new member of the MeV blazar family: PMN~J0641$-$0320.
Our optical spectroscopy { provides confirmation that} this is a flat--spectrum radio quasar located at a redshift of $z=1.196$.
{ Its} very hard $NuSTAR$ spectrum 
(power--law photon index of $\sim$1 up to $\sim$80 keV)
{ indicates that the emission  is produced via inverse Compton
  scattering off photons coming from outside the jet.}
The overall spectral energy distribution { of PMN J0641$-$0320} is typical of powerful blazars
{ and by reproducing it with a simple one-zone leptonic emission
  model we find the emission region to be located either inside the
  broad line region or within the dusty torus.}

%and argues for the high-energy emission being powered by inverse Compton
%scattering { off} the photons of the broad line region.
\end{abstract}

%%%%%%%%%%%%%%%%%%%%%%%%%%%%%%%%%%%%%%%%%%%%%%%%%%%%%%%%%%%%%%%%%%
\section{Introduction}

Blazars are an extreme class of active galactic nuclei (AGN) whose
bright and violently variable panchromatic emission is ascribed to the presence
of a collimated relativistic jet closely aligned to our line of sight \citep[e.g.][]{blandford78}. 
These objects are typically hosted in the nuclei of giant elliptical
galaxies \citep{falomo00,odowd02} and can be powered by { accretion onto} larger--than--average
super--massive black holes \citep[see e.g.][]{ghisellini10,shaw13}.
Blazars are sub-classified into flat-spectrum radio quasars (FSRQs) and BL
Lacertae (BL Lac) objects depending on the presence (or absence for BL
Lacs) of emission lines  in their optical
spectrum with equivalent width $>$5\,$\AA$ { \citep[e.g.,][]{urry95,marcha96}}.

Among all blazars, the so--called `MeV blazars', those having an inverse Compton peak
located in the MeV band \citep{bloemen95,sikora02,sambruna06}, may be the most extreme objects.  
These rare, extremely luminous objects are mostly found at high ($z>$2--3)
redshift and are thought to host super--massive black holes with
masses often in excess of 10$^9$\,M$_{\sun}$ \citep[e.g.][]{ghisellini10}.
{ Since} each detected blazar implies the presence\footnote{For
each detected blazar with a bulk Lorentz factor $\Gamma$, the total number
of objects with jets pointing in all directions is of the order of 2$\Gamma^2$.}
 of a much larger population
of objects with jets pointing somewhere else, the few detections of these extreme
blazars are instrumental to set robust constraints on the mass function of heavy
black holes. This becomes particularly important at redshift $z>$4 when
the age of the Universe is barely compatible with the time needed
to grow such monstrous black holes  exclusively
by accretion \citep{volonteri11,ghisellini13}. All this has sparked a renewed interest
in this elusive, yet interesting, class of blazars.

Lacking an MeV all--sky instrument, the most efficient domain in which
to detect MeV blazars is the hard X--ray ($>$10\,keV) band. In this
energy range, such objects display remarkably hard spectra, which  easily distinguish
them from other, more normal, sources. 
The {\it Swift} Burst Alert Telescope (BAT)
survey detected 26 flat--spectrum radio quasars (FSRQs) of which $\sim$40\,\%
are at $z>$2 \citep{ajello09} and host massive black holes \citep{ghisellini10}.
This is in  contrast to {\it Fermi}-LAT which has detected $>$400 FSRQs, but only
$\sim$12\,\% of those are located at $z>$2 \citep{3LAC}. 
This is mostly due
to the fact that high--redshift FSRQs are {\it soft} $\gamma$--ray sources (e.g.
power--law photon indices $>$ 2.4--2.5) and 
since the LAT point spread function increases at low energies, it is hard to disentangle point source emission from the bright diffuse Galactic emission.

In the absence of an all--sky hard X--ray survey more sensitive than the one obtained with 
{\it Swift}/BAT, MeV blazar candidates have recently been identified on the basis of radio, 
IR, optical and soft X--ray observations \citep[e.g.][]{sbarrato12,ghisellini14}
and then later confirmed by {\it NuSTAR} hard X--ray observations \citep{sbarrato13}.
Another strategy relies on the detection and identification 
of MeV blazars during flaring episodes at $\gamma$ rays.
Here we report on the {\it Fermi} detection  of the transient source
Fermi J0641$-$0317 \citep{kocevski14_atel} later identified to be coincident 
with the radio source PMN J0641$-$0320 \citep{ajello14_atel}.  
Because of its potentially interesting nature, we initiated a
multi-wavelength campaign and here we present the results of the target of opportunity (ToO)
observations carried out by {\it Swift} and {\it NuSTAR} in X--rays,
and  with GROND\footnote{GROND is an optical/NIR camera mounted 
  on the MPG 2.2m telescope in La Silla, Chile \citep{greiner08b}.
}, in the optical/NIR,
that firmly establishes PMN~J0641$-$0320 as a new member of the MeV blazar family.

%%%%%%%%%%%%%%%%%%%%%%%%%%%%%%%%%%%%%%%%%%%%%%%%%%%%%%%%%%%%%%%%%%
\section{Observations}
\label{sec:obs}

\subsection{Fermi}
\begin{figure*}[ht!]
  \begin{center}
  \begin{tabular}{c}
	 \includegraphics[scale=0.7,angle=0]{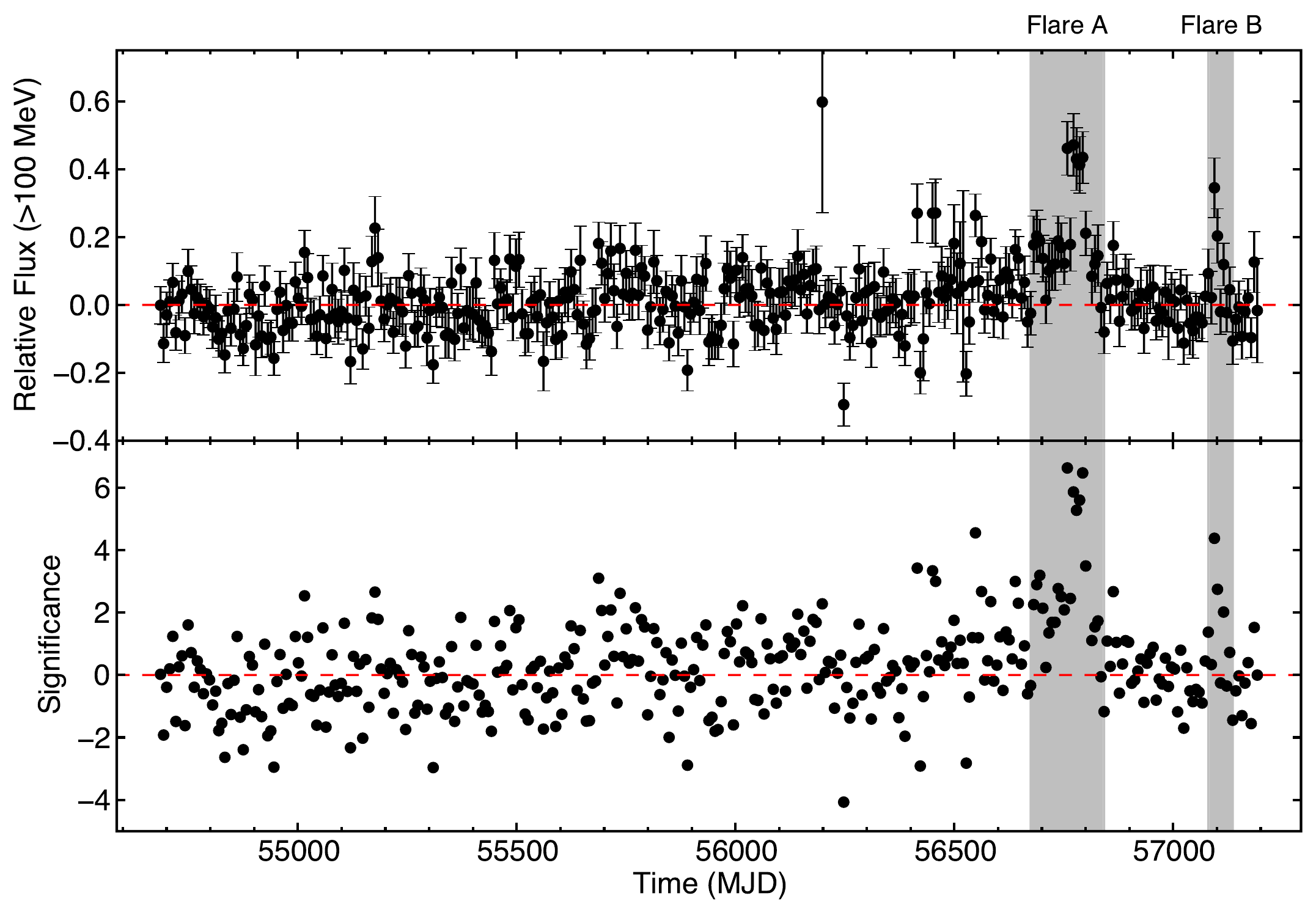}
\end{tabular}
  \end{center}
\caption{FAVA relative flux  (upper panel) and significance (lower panel)
light curve of Fermi J0641--0317. The relative flux is the excess flux
at that position \citep[i.e. the total flux minus the mission averaged flux
at that position, see e.g.][]{fava13} divided by the mission averaged flux, while the significance is
expressed in units of the standard deviation ($\sigma$) of a Gaussian normal distribution.
Note the significant flux increase  in the week of 2014  April 14--21 (around
MJD 56760), which is part of a longer flare indicated as `Flare A' and marked in gray.
There is also a second flare (`Flare B') around MJD 57100 (see text).
\label{fig:fava}}
\end{figure*}

Fermi J0641$-$0317 was detected as a significant ($>$6\,$\sigma$)  
$\gamma$--ray transient during 2014 April 14--21 week \citep[and
reported to the community in an {Astronomer's Telegram,}][]{kocevski14_atel}
by the {\it Fermi} all--sky variability analysis \cite[FAVA,][]{fava13}. 
FAVA is a real-time analysis
that searches the $\gamma$--ray sky for weekly transients and detects 
significant deviations above the mission-averaged flux at every
position in the sky. FAVA is an efficient tool to detect weekly
transient all over the sky.
Figure~\ref{fig:fava} reports the FAVA light--curve of  Fermi J0641$-$0317 with
the $>$6\,$\sigma$ flaring episode detected around MJD 56800.

% ----------------------------------------------------------
% ----------------------------------------------------------

Using the standard {\it Fermi} science tools\footnote{http://fermi.gsfc.nasa.gov/ssc/data/analysis/}  
and {\tt P7SOURCE} photons,
the position of Fermi J0641--0317 was reported  to be \citep[for 
2014 April 14--21 week, see ][]{kocevski14_atel} at
R.A. = 100.383 deg, Dec. = --3.294 deg (J2000) with a 95\,\% confidence region
of 0.25\,deg. %Subsequent {\it Swift} observations (described later) found within the \lat error circle a bright X-ray transient, positionally coincident with PMN J0641-0320, a bright flat-spectrum radio source \citep{ajello2014_atel}.
This source is located in the plane of the Galaxy ($b=-3.703$),
but towards the anti--center region.
Its $\gamma$--ray spectrum, { covering the period 2014 April 14--21 and} modeled with a power law, exhibited a 
0.1--300\,GeV flux of $(7.7\pm 1.3)\times10^{-7}$ ph cm$^{-2}$ s$^{-1}$
and a photon index of $2.66\pm 0.15$. 
For comparison, less than 10\% of the 
{\it Fermi}--LAT detected FSRQs have a larger photon index \citep{2LAC}.
Thus, even during the flare, Fermi J0641$-$0317 displayed a very soft $\gamma$--ray spectrum.

A source coincident with Fermi J0641$-$0317 was later reported (as 3FGL~J0641.8$-$0319) 
in the 3FGL catalog \citep{3FGL} based on four years of {\it Fermi}--LAT observations. 
Its $>$100 MeV
{\
flux averaged over four years was 
% $>$100 MeV flux of 
($1.7\pm 0.5)\times 10^{-8}$ ph cm$^{-2}$ s$^{-1}$,  
about 45 times fainter than during the flare.
The power--law photon index was $2.45\pm0.13$,
similar (within the uncertainties) to the slope during the flare.
}
% are, respectively,  fainter and harder of the values recorded during the flare.
The long--term FAVA light--curve confirms that the source had, over the course of {\it Fermi}--LAT
observations, a total of two flaring episodes: between 2014 January 1 and 2014 July 4 and 
between 2015 February 27 and 2015 April 24
{
(see Figure~\ref{fig:fava}).
}
Here we take advantage of the newly delivered Pass~8 dataset to re--analyze the data of Fermi J0641$-$0317. 
We use  {\tt P8\_SOURCE} photons, the {\tt P8R2\_SOURCE\_V6} instrument response function and rely on 
version {\tt 10-00-04} of the {\it Fermi} science tools.
The analysis was performed, following the 
recommendation\footnote{http://fermi.gsfc.nasa.gov/ssc/data/analysis/documentation/Cicerone/Cicerone\_Data\_Exploration/Data\_preparation.html} 
for the analysis of a point source in the plane of the Galaxy, in a region of interest 
(ROI) centered on the source and with a radius of  15$^{\circ}$. 
All photons detected at zenith angles  larger than 90$^{\circ}$ were removed.
The background model comprised the diffuse Galactic and isotropic
emission and all 3FGL sources  \citep{3FGL} 
within 20$^{\circ}$ of the source. { The spectral parameters of all
  the sources present within the ROI 
  were left free to vary during the likelihood fitting.}

\begin{figure*}[!ht]
\begin{center}
\begin{tabular}{lc}
\hspace{-1cm}
    \includegraphics[scale=0.47]{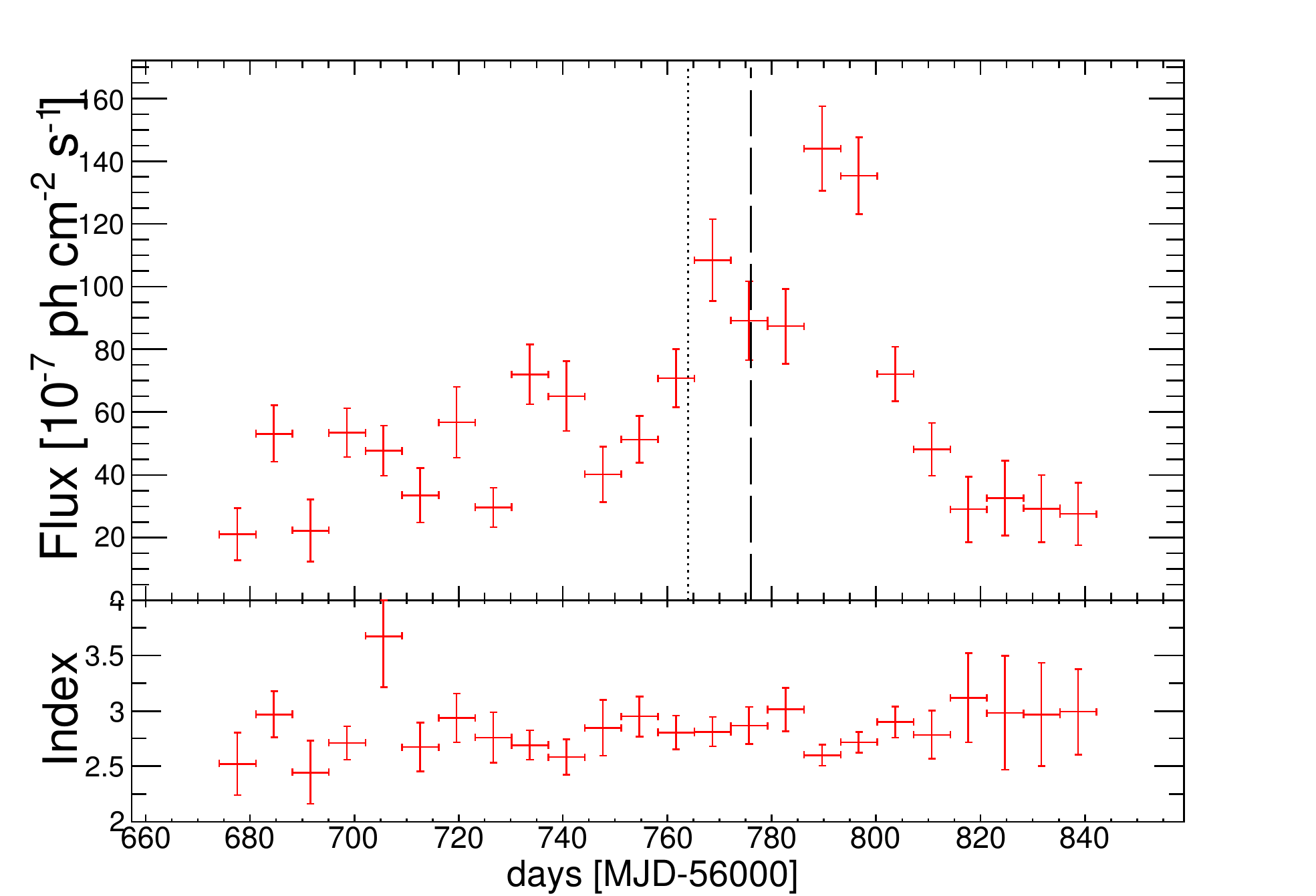} &
\hspace{-1cm}
    \includegraphics[scale=0.47]{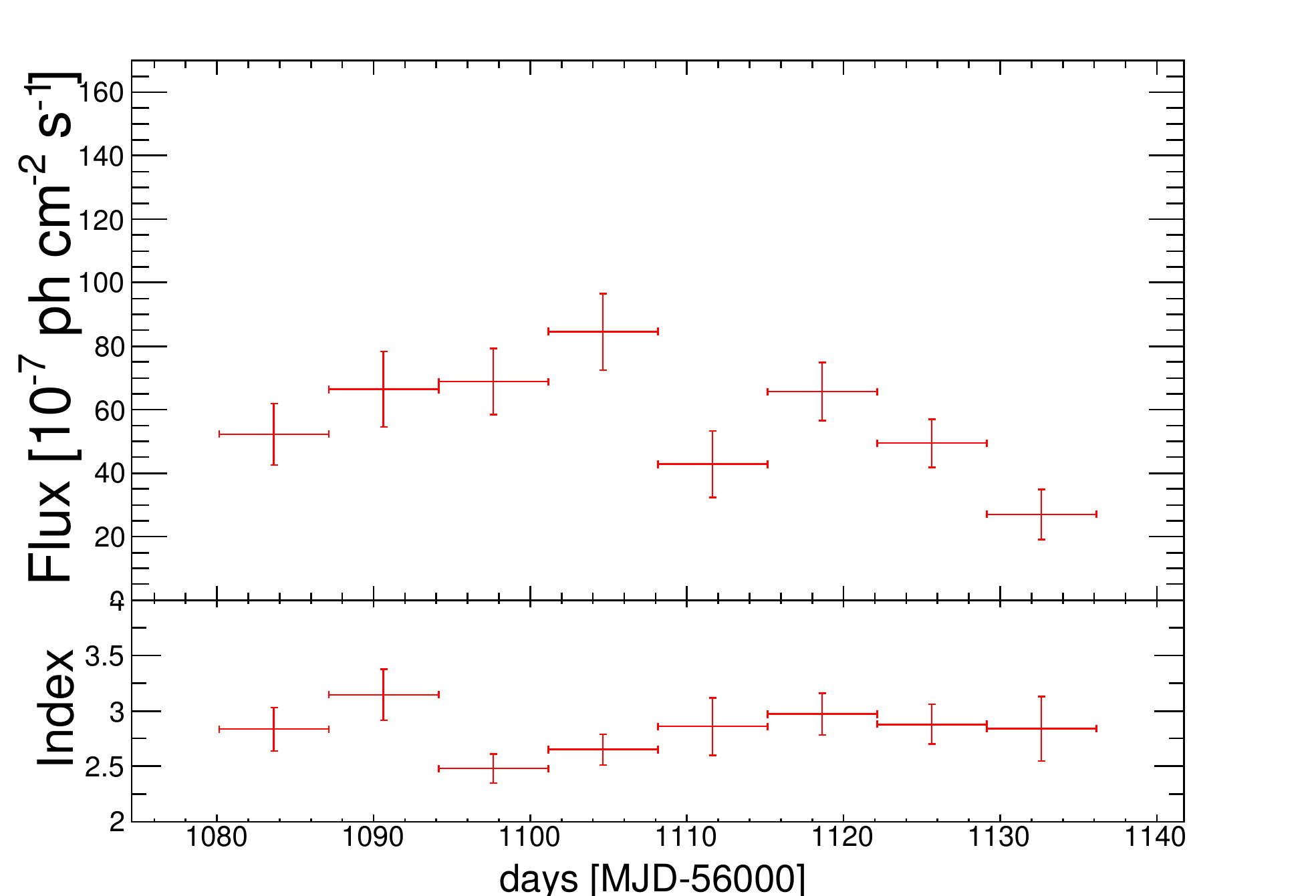} 
\end{tabular}
\end{center}
 \caption{Maximum likelihood ($>$100\,MeV) weekly light curves of flare A (left) and flare B (right).
 The short dashed and long dashed lines show the times when \lat\ first detected the source and when {\it NuSTAR} observed it.
    \label{fig:lat_flares}}
\end{figure*}

During both long--term flares the source is well detected (with a test
{ statistic}, TS\footnote{The significance of each source is evaluated using the test statistic  
${\rm TS}=2(\ln \mathcal{L}_1 - \ln \mathcal{L}_0)$, where $\mathcal{L}_0$ and $\mathcal{L}_1$
are the likelihoods of the background (null hypothesis) and 
the hypothesis being tested (e.g. source plus background). The
significance of the detection can be expressed in terms of the number of
standard deviation of a normal Gaussian distribution as $n_{\sigma}\approx\sqrt{T}$. }, of 1591 and 471 respectively) and with similar spectral 
parameters (see Table~\ref{tab:results}).
The weekly light--curves for both flares (reported in Figure~\ref{fig:lat_flares}) show that the source was 
significantly detected by the LAT, with several flaring episodes approaching fluxes ($>$100\,MeV) of 
10$^{-6}$\,ph cm$^{-2}$ s$^{-1}$ accompanied, during those times,
by a slightly harder than average spectrum.
Figure~\ref{fig:lat_flares} shows that FAVA first detected the source during the main flare
and that the source was still bright during the {\it NuSTAR} observation. However, the source reached
its maximum two weeks later (on 2014 May 13) reaching a flux of $(1.44\pm0.13)\times10^{-6}$\,ph cm$^{-2}$ s$^{-1}$
with a power--law photon index of 2.60$\pm0.10$. 
%At the maximum the spectrum was harder than the average spectrum detected during both flares that exhibited a photon index of $\approx$3. { Non capisco, non era 2.66??? }

%%%%%%%%%%%%%%%%%%%%%%%%%%%%%%%%%%%%
%
% SWIFT
%
%%%%%%%%%%%%%%%%%%%%%%%%%%%%%%%%%%%%%

\subsection{{\it Swift} Observations}

\begin{figure*}[ht!]
  \begin{center}
  \begin{tabular}{c}
	 %\includegraphics[scale=0.82]{Images/fermi_xrt_image.eps}
%\includegraphics[scale=0.6]{Material/xrt_map.pdf} \\
%\vspace{-15cm}
\includegraphics[scale=0.6,clip=true,trim=0 170 0
    160]{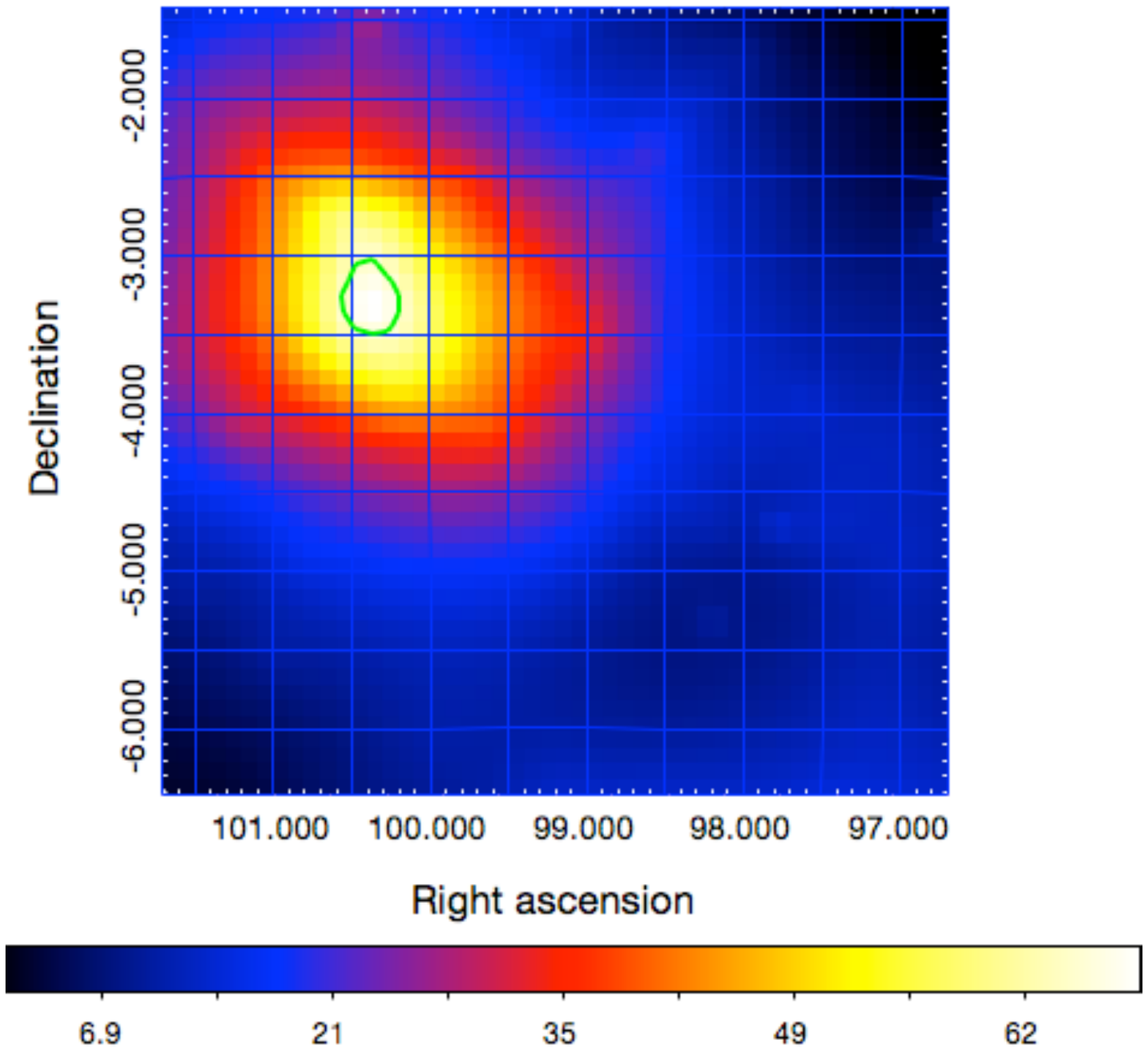}\\
\vspace{-0.5cm}
\includegraphics[scale=0.5,clip=true,trim=0 100 0 150]{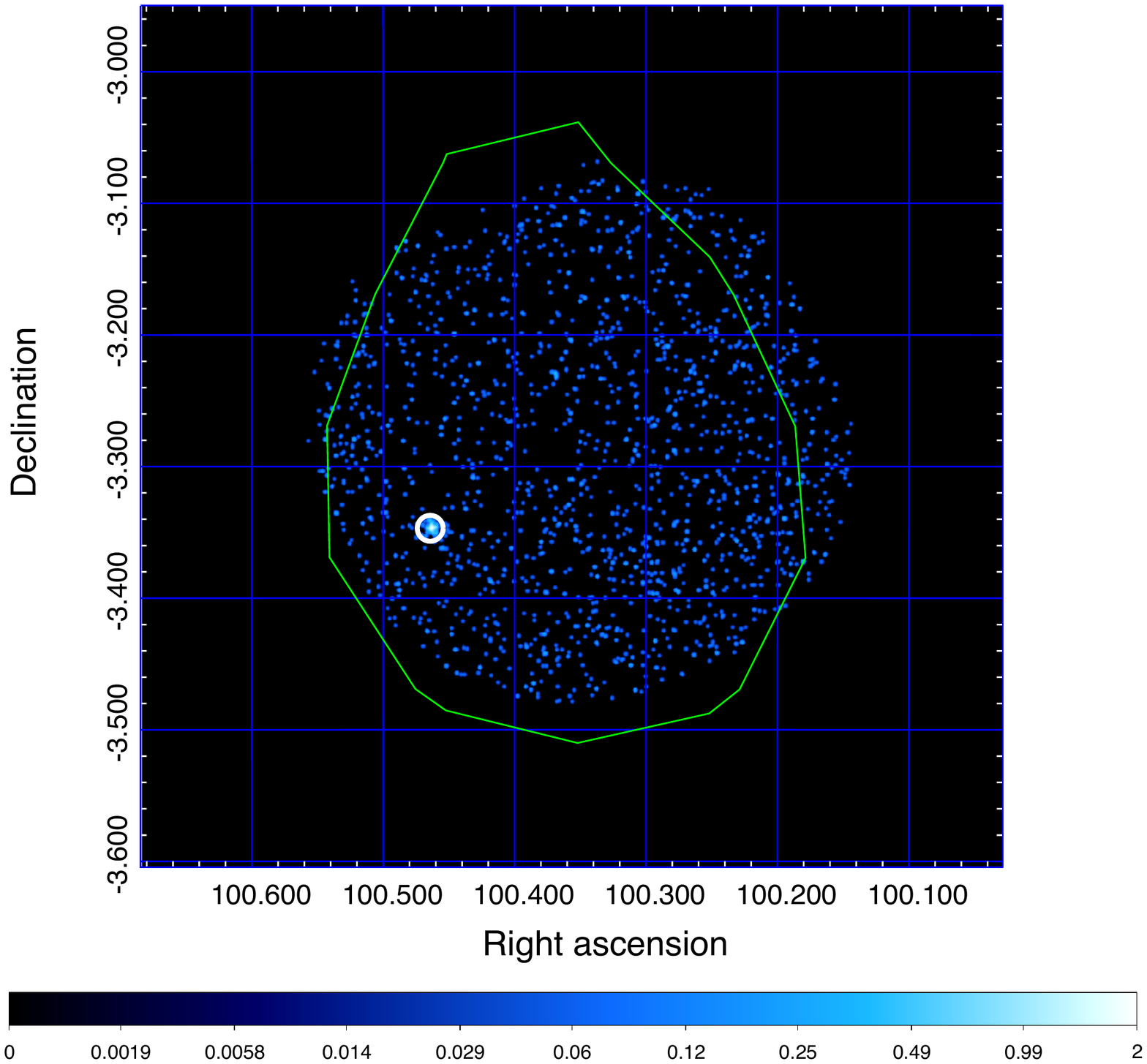}
\end{tabular}
  \end{center}
\vspace{-0.5cm}
\caption{{ Top Panel}: {\it Fermi}--LAT test statistic (TS) map at the position
of the transient Fermi J0641--0317 for the week of April 14--21
(2014). The map shows at every pixel the likelihood (in term of TS as
color coded in the color bar) of the source being at that pixe.
The green contour shows the 95\% error region on the position of the source. 
{ Bottom Panel}: {\it Swift}--XRT observation of April 26th with super--imposed
the 95\% {\it Fermi}--LAT error region. 
The white circle shows the position of the only source detected. 
The X-ray source position coincides  with that of the known radio
source PMN~J0641$-$0320.
The image was smoothed and the color { bars} show the number of counts per pixel.
\label{fig:tsmap}}
\end{figure*}

{\it Swift} was triggered to perform three observations: on April 24th, April 26th and April 29th 2014,
{ the last} one happened simultaneously to the {\em NuSTAR} observation (see later).
In all the epochs, only one bright source was detected by the X--ray telescope (XRT)
within the error region of the LAT. 
The source was localized to R.A.= 6h41m51.20s, Dec.= --3:20:46.34  (J2000)
with a 90\,\%  uncertainty of 3.7\,arcsec. Figure \ref{fig:tsmap} shows the {\it Fermi}--LAT
and {\it Swift}--XRT localizations. 
In all observations the source remained
very bright with a 2--10\,keV flux of $\gtrsim 5\times10^{-12}$ erg cm$^{-2}$ s$^{-1}$ 
and displayed a very hard spectrum with a photon index of
$\sim$1.0. { As can be seen in Table \ref{tab:results}, in the {\it
    Swift}-XRT observations, 
there is a marginal ($\leq$2\,$\sigma$) evidence for variability from one pointing to another, 
at the level of $\sim$40\%.}
The source flux extrapolated to the 15--150 keV band is 
$\sim10^{-10}$ erg cm$^{-2}$ s$^{-1}$, which would make it easily
detectable by the {\it Swift} Burst Alert Telescope (BAT) in less than 10$^5$\,s 
\citep[see e.g.][]{ajello08a,tueller08}. The lack of such { a} source in the most 
recent BAT catalogs that rely on $>$50\,months of exposure 
\citep{cusumano10,ajello12,baumgartner13} testifies
that this might be an unusually high/hard X--ray state for this source.

% ----------------------------------------------------------
% ----------------------------------------------------------

The prominent flat--spectrum radio source PMN J0641$-$0320 \citep{fomalont03} lies only
2.4 arcsec away from the XRT centroid and well within its error radius. 
The source was resolved in prior very long baseline array (VLBA) observations
at 8.6\,GHz showing parsec--scale emission and a total flux of 0.83\,Jy.
The radio brightness and the small angular separation between the XRT and the radio
source make the probability that the radio source is a background unrelated
object negligible \citep{petrov13}. 
Moreover, within the error box of XRT and compatible with the radio position of PMN J0641$-$0320  
there is a source detected at infrared by WISE \citep[J064151.12-032048.4,][]{wright10}
with IR colors typical of blazars
\citep{massaro2011,dabrusco12,dabrusco2014}.
The source also has  a flat spectrum below 1\,GHz { which is typical} for
$\gamma$-ray detected blazars \citep{massaro2013a,massaro2013b}.
 We thus consider the association
of the blazar--like source PMN~J0641$-$0320 to the transient Fermi J0641$-$0317 very robust.

%%%%%%%%%%%%%%%%%%%%%%%%%%%%%%%%%%%%
%
% NuSTAR
%
%%%%%%%%%%%%%%%%%%%%%%%%%%%%%%%%%%%%%

\subsection{NuSTAR Observations}

%NuSTAR carried out a 20\,ks observation on 2014 April 29  simultaneously to one of the {\it Swift} observations. PMN J0641--0320 is detected significantly by both focal plane modules (referred here as FPMA and FPMB) with a total of $\sim$6000 background--subtracted counts.

PMN~J0641$-$0320 was observed with {\em NuSTAR} \citep{harrison13} starting at UT 10:01 on 2014 April 29 (MJD~56776). 
The target was observed for 11 hours, resulting in 21.4~ks of source exposure after event filtering. 
Data were processed using the {\em NuSTAR} Data Analysis Software (NuSTARDAS; \citealt{perri13}) v.1.2.1, 
and response files from v.~2013090 of the Calibration Database.
% NOTE: This is not the latest version, but before the paper is submitted I will verify that we 
% obtain consistent results with the most recent version. For simple observations like this one, nothing is expected to change.
We extracted the {\em NuSTAR} source and background spectra from filtered event files using the standard 
{\em nuproducts} script. 
For the source we used circular extraction regions with a diameter of
60\arcsec\ for both focal plane modules (referred to as FPMA and FPMB). 
The background was extracted from large annular regions centered on the source. 
The choice of extraction region size optimizes the signal--to--noise 
ratio at high energies; we have verified that alternative choices do not affect any of the results. 
Due to the very hard spectrum, the target is well detected up to the high-energy end of the {\em NuSTAR} 
bandpass at $\sim$70~keV. 
No variability is apparent within the {\em NuSTAR} observation.

For spectral modeling, we bin the {\em NuSTAR} spectra to a minimum of 20 counts per bin.
% NOTE: Alternatively, we could also use minimum SNR per bin and no results would change. 
% It depends on what is used for joint NuSTAR and Swift/XRT modeling, to keep things self-consistent.
We use {\em Xspec} v.~12.8.1 \citep{arnaud96}, and a simple power-law
model $dN/dE\propto E^{-\Gamma_X}$ for the photon spectrum.
The neutral hydrogen column density in the direction of PMN~J0641$-$0320 of $6\times10^{21}$~cm$^{-2}$ 
\citep{kalberla05} is too low to significantly attenuate the spectrum above 3~keV, but we include a 
fixed absorption factor for completeness. 
The best fit is obtained for a very hard photon index $\Gamma_X=1.08\pm0.03$ (90\% confidence interval), 
with no structure apparent in the residuals and $\chi^2=285$ for 275 degrees of freedom. 
The cross--normalization constant between FPMA and FPMB was left free to vary in the fit and found 
to be $1.02\pm0.04$, consistent with expectations from  calibration
observations \citep{madsen2015}. 
The flux calculated from the power-law model is 
$(6.8\pm0.2)\times10^{-12}$~erg\,s$^{-1}$\,cm$^{-2}$ for the 2--10~keV
energy band and 
$(4.5\pm0.2)\times10^{-12}$~erg\,s$^{-1}$\,cm$^{-2}$ for the 10--70\,keV band. 
We place an upper limit on the curvature within the {\em NuSTAR} bandpass by 
fitting a log--parabolic model \citep{tramacere2007} with one additional parameter ($f(E)\propto E^{-\alpha_x-\beta_x \log E}$), 
which leads to $\beta_x<0.09$ with 90\% confidence.

\begin{figure*}[ht!]
  \vskip -0.5 cm
  \begin{center}
	 \includegraphics[scale=0.82]{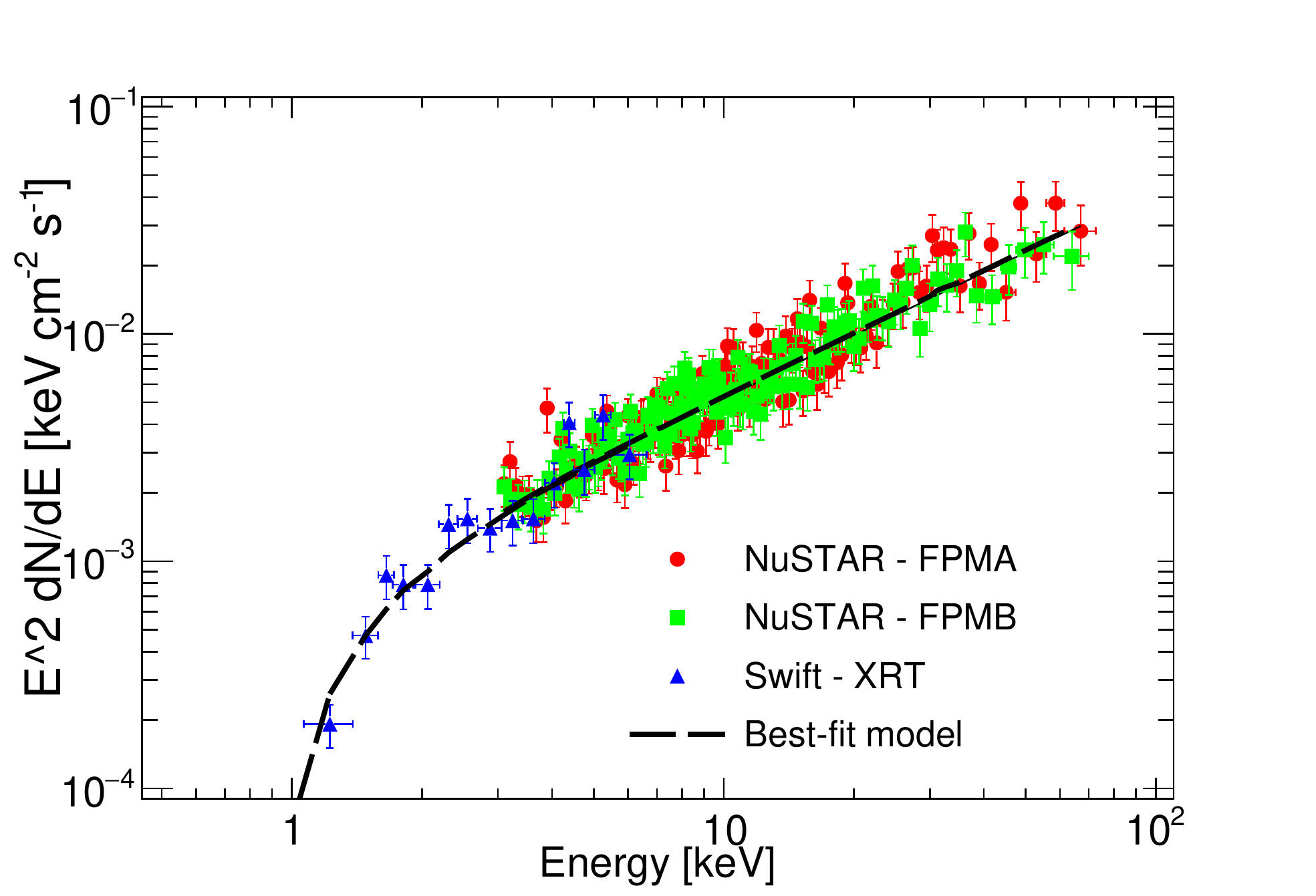}
  \end{center}
  \vskip -0.5 cm
\caption{{\it NuSTAR} and {\it Swift}/XRT observation of PMN J0641--0320 on 2014 April 29.
The dashed line is the best-fitting absorbed power-law model described in the text. 
The absorption is compatible with the Galactic absorption along the line of sight.
\label{fig:nustar}}
\end{figure*}

Figure~\ref{fig:nustar} shows the joint fit to the {\it
  Swift}-XRT/{\em NuSTAR} datasets for the simultaneous observation
performed on April 29. It is apparent that the two observations are in
agreement with each other and that, over the entire 1--70\,keV energy range,
the spectrum of the source can be described (see Table~\ref{tab:results}) as
a simple (very hard) power law with a photon index of 1.06$\pm0.03$
absorbed by Galactic gas and dust along the line of sight \citep{kalberla05}.

%%%%%%%%%%%%%%%%%%%%%%%%%%%%%%%%%%%%%%%%%%%%%%%%%%%%%%%%%%%%%%%%%%
\begin{deluxetable}{lccccc}
\tablewidth{0pt}
\tabletypesize{\scriptsize}
\tablecaption{Table of Observations and Spectral Parameters
\label{tab:results}}
\tablehead{\colhead{Instrument}  & \colhead{Date\tablenotemark{a}} &
\colhead{Energy Band} & \colhead{Flux\tablenotemark{b}}  & 
\colhead{Photon Index\tablenotemark{c}} &
\colhead{Description}
}
\startdata
% -------------------------------------------------------------------------------
%%%% LAT
LAT & 04/14--04/21 & 0.1--500\,GeV & (7.7$\pm$1.3)$\times10^{-7}$ & 2.66$\pm0.15$ & LAT First detection\\
LAT\tablenotemark{d} & 04/24--05/01 & 0.1--500\,GeV &
(8.2$\pm$0.1)$\times10^{-7}$ & 2.68$\pm0.15$& {\it NuSTAR} observation\\
LAT & 01/17--07/04 & 0.1--500\,GeV & (5.9$\pm$0.2)$\times10^{-7}$ & 2.79$\pm0.03$ & Flare A \\
LAT & (2015) 02/27--04/24 & 0.1--500\,GeV &
(5.5$\pm$0.1)$\times10^{-7}$ & 2.80$\pm0.03$ & Flare B\\
\hline
%%%% XRT
XRT & 04/24  & 2--10\,keV  & 4.9$^{+1.0}_{-0.5}\times10^{-12}$ & 0.93$^{+0.43}_{-0.39}$\\
XRT & 04/26  & 2--10\,keV  & 6.2$^{+0.9}_{-1.3}\times10^{-12}$ & 0.87$^{+0.43}_{-0.35}$\\
XRT\tablenotemark{d} & 04/29  & 2--10\,keV  & 7.2$^{+0.9}_{-1.3}\times10^{-12}$ & 0.93$^{+0.33}_{-0.30}$\\
\hline
%%%% NuSTAR
{\em NuSTAR}\tablenotemark{d} & 04/29  & 3--70\,keV  & 5.2$^{+0.3}_{-0.3}\times10^{-12}$ & 1.08$^{+0.03}_{-0.03}$\\
\hline
XRT+{\em NuSTAR}\tablenotemark{e} & 04/20 & 1-70\,keV  &
8.9$^{+0.2}_{-0.2}\times10^{-12}$ & 1.06$^{+0.03}_{-0.03}$\\

\enddata
\tablenotetext{a}{
All dates of observations are in 2014 unless otherwise noted.}
\tablenotetext{b}{{\it Fermi}-LAT fluxes are in ph cm$^{-2}$ s$^{-1}$, 
{\it Swift}--XRT and {\it NuSTAR} fluxes are in erg cm$^{-2}$ s$^{-1}$.}
\tablenotetext{c}{Photon index of the power-law model fitted to the data.}
\tablenotetext{d}{Data used for building the simultaneous SED reported
  in Figure~\ref{sed}.}
\tablenotetext{e}{These are the results of the joint-fit to XRT and
  {\em NuSTAR} data for the simultaneous observations on 04/29.}
\end{deluxetable}
% -------------------------------------------------------------------------------

% -------------------------------------------------------------------------------
% -------------------------------------------------------------------------------

%---------------------------------
\begin{table*} 
\centering
\footnotesize
\begin{tabular}{llllllll}
\hline
\hline
~  &$g^\prime$ &$r^\prime$ &$i^\prime$ &$z^\prime$ &$J$ &$H$ &$K_{s}$ \\
\hline   
$\lambda_{\rm eff}$ (\AA)   &4587 &6220 &7641 &8999 &12399 &16468 &21706 \\  
%AB mag      &... &22.16$\pm$0.16 &19.91$\pm$0.04 &19.73$\pm$0.03 &19.52$\pm$0.06 &19.20$\pm$0.09 &19.35$\pm$0.24 \\  
mag$_{\rm AB}$ & $22.26\pm0.25$ & $20.76\pm0.08$ & $20.01\pm0.08$ &
$19.32\pm0.05$ & $18.40\pm0.11$ & $18.83\pm0.12$ & $17.09\pm0.20$ \\
\hline
\hline 
\end{tabular}
\vskip 0.4 true cm
\caption{GROND AB observed magnitudes of PMN J0641--0320, taken UT 2014 April 25  
(magnitudes not corrected for Galactic foreground extinction). 
The first row gives the effective wavelength of the filter (in Angstroms).
}
\normalsize
\label{grond}
\end{table*}
%---------------------------------

\subsection{GROND observations}

On 2014 April 26 01:01 UTC  PMN~J0641$-$0320 was observed simultaneously
in four optical ($g^\prime$′, $r^\prime$, $i^\prime$, $z^\prime$) and three
NIR ($J$,$H$,$K$) bands with the GROND \citep{greiner08b} instrument at the 2.2\,m
MPG telescope at  La Silla Observatory (Chile). 
Single exposures were obtained with 142\,s integrations in the optical bands and
240\,s integrations in the NIR bands. Observing conditions were moderate
with a seeing of $1.8^{\prime\prime}$ and an average airmass of 2.0.

Data reduction and photometry were performed using standard IRAF tasks
\citep{tody93},  similar to  the  procedure  described in  \citep{kruehler08}.  
The  $g^{\prime},  r^{\prime},  i^{\prime},  z^{\prime}$
photometry  was  obtained  using point-spread-function  (PSF)  fitting
while due  to the under--sampled  PSF in  the NIR, the  $J, H,
K_s$ photometry  was measured  from apertures  with  sizes
corresponding to the Full--Width at Half Maximum (FWHM) of field stars. 

The optical photometry was calibrated against an SDSS--field calibrated
observation of the same field taken { on} a different night under photometric conditions.
Photometric calibration of the NIR bands was achieved  against
selected  2MASS  stars \citep{skrutskie06}
in the field of the blazar.

The resulting AB magnitudes, not corrected for the predicted Galactic
foreground reddening of $E_{\rm B-V}=0.98$\,mag \citep{schlafly11},
are presented in Table~\ref{grond}.

%%%%%%%%%%%%%%%%%%%%%%%%%%%%%%%%%%%%%%%%
%
%  Keck spectroscopy subsection
%
%%%%%%%%%%%%%%%%%%%%%%%%%%%%%%%%%%%%%%%%

% -------------------------------------------------------------------------------
\begin{figure*}[ht!]
% \vskip -0.5 cm
  \begin{center}
	 \includegraphics[scale=0.5,angle=-90]{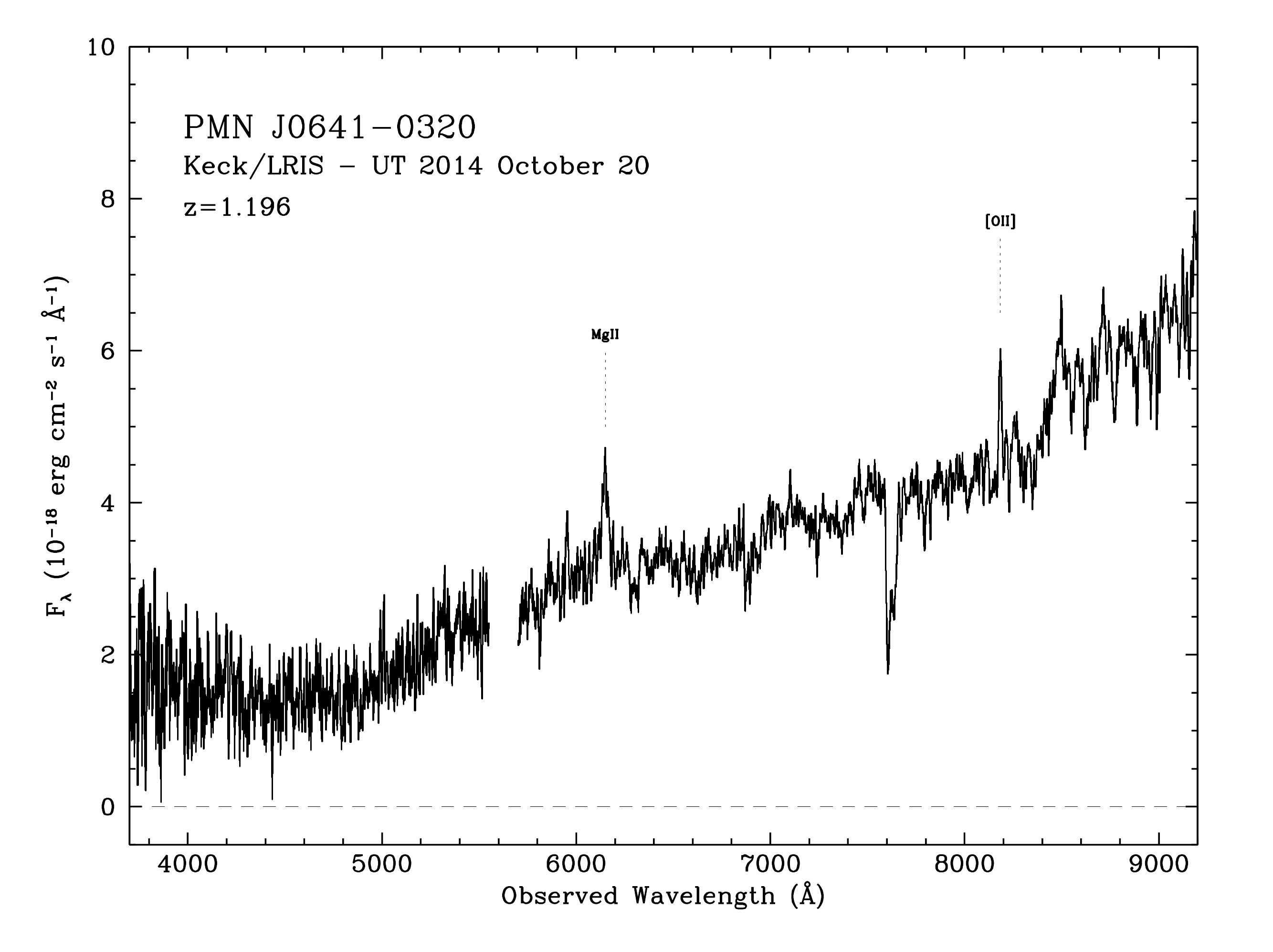}
  \end{center}
  \vskip -0.5 cm
\caption{Optical spectrum, acquired with Keck, of PMN~J0641--0320
\label{fig:keck}}
\end{figure*}
% -------------------------------------------------------------------------------

\subsection{Keck Spectrum}

We obtained an optical spectrum of PMN~J0641-0320 on UT 2014 October
20 using the Low Resolution Imaging Spectrometer \citep[LRIS;][]{oke95},
a dual-beam spectrograph on the Keck~I telescope atop Mauna Kea.
The conditions were poor due to Hurricane Ana, with significant
clouds.  We observed the target through a 1\farcs0 slit for two
600~s exposures using the 600 $\ell\, {\rm mm}^{-1}$ grism on the
blue arm of the spectrograph ($\lambda_{\rm blaze} = 4000$~\AA,
resolving power $R \equiv \lambda / \Delta \lambda \sim 1000$), the 
400 $\ell\, {\rm mm}^{-1}$ grating on the red arm of the spectrograph
($\lambda_{\rm blaze} = 8500$~\AA, $R \sim 1200$), and the 5600~\AA\
dichroic.  The data were processed using standard techniques within
IRAF, and because no standard stars were taken on that cloudy night,
we flux calibrated the spectrum using an archival sensitivity
function with the same instrument configuration.

The optical spectrum (displayed in Figure~\ref{fig:keck}) shows strong, red continuum with two emission
lines that we identify as broad \ion{Mg}{2}~$\lambda 2800$ and
narrow [\ion{O}{2}]~$\lambda 3727$.  The broad line has an observed 
equivalent width of $\sim 15$~\AA\ and a full-width at half maximum
of FWHM $\sim 2000\, {\rm km}\, {\rm s}^{-1}$, clearly indicating
a quasar.

The \ion{Mg}{2}~$\lambda 2800$ line was confirmed a few { nights}
later  using Magellan.
Our spectroscopic observations thus place the object at a redshift of
$z=1.196$. Because of its optical and radio properties, PMN~J0641-0320
is a new flat-spectrum radio quasar.

% -------------------------------------------------------------------------------
\begin{figure*}[ht!]
  \begin{center}
\vskip -1 cm
\includegraphics[scale=0.75]{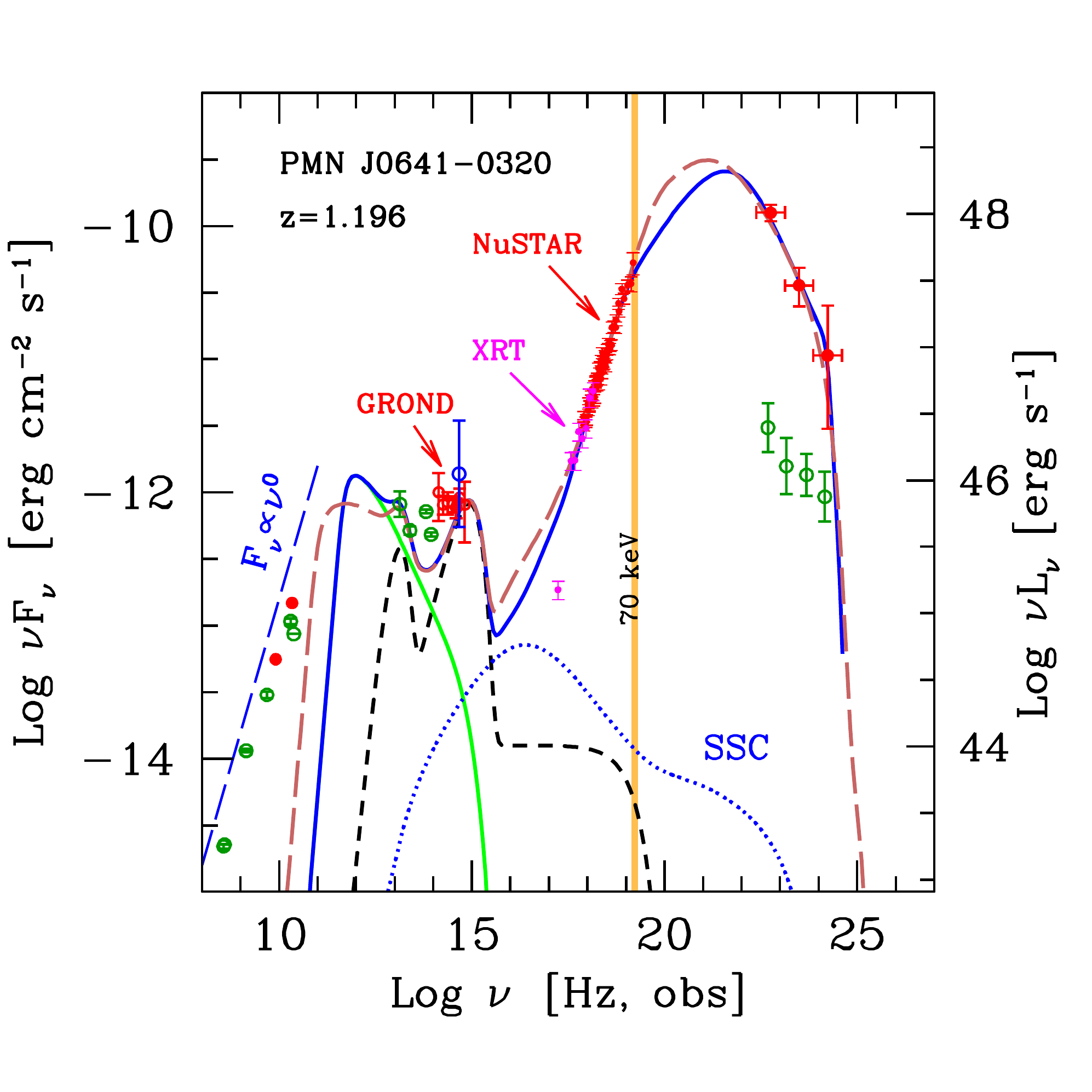}
\end{center}
  \vskip -1.5 cm
\caption{
Overall SED of PMN~0641--0320 together with the one--zone leptonic
model we have used to interpret the SED.
Red circles correspond to quasi--simultaneous data, green symbols are archival data.
% The blue line corresponds to a redshift $z=1$, the light blue line to $z=2$.
The black short dashed lines correspond to the contribution from the IR torus,
the accretion disk and the X--ray corona.
{
The solid blue line corresponds to a dissipation region lying within the BLR,
while the long dashed brown line corresponds to $R_{\rm BLR}< R_{\rm diss} <R_{\rm torus}$.
The solid (green) lines correspond to the synchrotron flux of the ``BLR" model.
The dotted blue line corresponds to the SSC emission for the same model.
% the dashed grey lines correspond to the synchrotron self Compton component.
}
The 3FGL spectrum is also reported.
\label{sed}}
\end{figure*}
% ----------------------------------------------------------------------------
%-----------------------------------------------------------------------------
\begin{table*}
\footnotesize
\centering
\begin{tabular}{lllllll lllll lllll }
\hline
\hline
Model &$M$  &$\Gamma$  &$R_{\rm diss}$  &$L_{\rm d}$  &$P^\prime_{\rm i}$ &$B$ &$\gamma_{\rm b}$ &$\gamma_{\rm max}$ &$s_1$  &$s_2$
&$\log P_{\rm r}$ &$\log P_{\rm B}$ &$\log P_{\rm e}$ &$\log P_{\rm p}$ \\
~[1]       &[2] &[3] &[4] &[5] &[6] &[7] &[8] &[9] &[10] &[11] &[12] &[13] &[14] &[15]\\ 
\hline
BLR   &1.1e9 &14  &240   &6.5 &0.07 &1.25 &170 &4e3 &0.5 &3.3  &46.6 &44.8 &45.2 &47.6  \\
Torus &1.1e9 &17  &1.2e3 &6.5 &0.12 &0.07 &1e3 &2e4 &1   &3.1  &46.9 &43.9 &46.0 &47.6 \\
% ~2 &13  &627  &33   &0.13  &1.17  &200 &4e3 &0.2 &4.1  &46.81 &45.53 &45.42 &47.68  \\
\hline
\hline
\end{tabular}
\vskip 0.4 true cm
\caption{
\small
{ Parameters of the model shown in Figure~\ref{sed}.
The two rows correspond to two locations of the dissipation region:
the first is within the BLR, the second is outside it, but within the torus.
For Figure \ref{sedcom} we use, an an illustration, the parameters of the first row. 
The spectral shape of the corona is assumed to be $\propto \nu^{-1} \exp(-h\nu/150~{\rm keV})$.
The X--ray corona emits 20\% { of} the disk luminosity.
We have assumed a viewing angle $\theta_{\rm v}=3^\circ$.
For $\Gamma=14$ (17), this implies $\delta=18.2$ (19).
Since we assume a conical jet of semi--aperture angle $\psi=0.1$ rad, the
size of the (assumed spherical) region is $R=\psi R_{\rm diss}$.
Thus $R=2.4\times 10^{16}$ cm for the ``BLR" case,
corresponding to a minimum observed variability timescale 
$t^{\rm obs}_{\rm var}= R (1+z) /(\delta c) \sim 27$ h.
For the ``torus" case, $R=1.2\times 10^{17}$ cm,
corresponding to  $t^{\rm obs}_{\rm var}=128$ h $= 5.4$ days.
} The columns are as  follows:
Col. [1]: model;
Col. [2]: black hole mass in solar units;
Col. [3]: bulk Lorentz factor;
Col. [4]: distance of the blob from the black hole in units of $10^{15}$ cm;
Col. [5]: disk luminosity in units of $10^{45}$ erg s$^{-1}$. 
{ 
The radius of the BLR is assumed to be
$R_{\rm BLR} =10^{17}L_{\rm d, 45}^{1/2}=2.6\times 10^{17}$ cm,
while the size of the torus is assumed to be 
$R_{\rm torus}=2.5\times 10^{18}L_{\rm d, 45}^{1/2}=6.4\times 10^{18}$ cm.
}
Col. [6]: power injected in the blob calculated in the comoving frame, in units of $10^{45}$ erg s$^{-1}$;
Col. [7]: magnetic field in Gauss;
Col. [8], [9]: { break} and maximum random Lorentz factors of the injected electrons;
Col. [10] and [11]: slopes of the injected electron distribution $Q(\gamma)$ below
and above $\gamma_{\rm b}$;
Col. [12] logarithm of the jet power in the form of radiation, [13] Poynting flux, [14]
bulk motion of electrons and [15] protons (assuming one cold proton
per emitting electron).
}
\label{para}
\end{table*}
%-----------------------------------------------------------------------------

\section{SED and Modeling}

Figure \ref{sed} shows the overall SED of PMN J0641--0320, together
with a fitted model.
The {\it Swift}--XRT and {\it NuSTAR} data are strictly simultaneous, 
while GROND and radio data \citep[provided by the RATAN-600, ][]{atel_ratan}
are quasi--simultaneous.
The {\it Fermi}--LAT data corresponds to a week integration time
centered 
(i.e. 3.5 days before and 3.5 days after) on the {\it NuSTAR} pointing 
(see Table \ref{tab:results}).
The other data are archival (green symbols).
% The scale of the right vertical axis corresponds to a redshift $z=1$.
% while the two shown models  correspond to $z=1$ and $z=2$.

% The unknown redshift of the source is the main obstacle for the modeling.
% However, the observed SED strongly suggests that that the source within the
% two considered redshifts.
% This is based on other powerful blazars observed with {\it Fermi}--LAT and {\it Swift}--BAT,
% showing a very strong and hard X--ray emission and a $\sim$MeV peak, implying
% a steep and relatively faint $\gamma$--ray flux above 100 MeV.

The adopted model is described in \cite{ghisellini09c}.
It is a one--zone, homogeneous leptonic model,
where the emitting particle distribution is derived through a continuity equation,
accounting for { continuous} injection, radiative cooling, and 
electron--positron pair production.
{
The resulting energy distribution of the emitting particles $N(\gamma)$ [cm$^{-3}$] 
is calculated after one light crossing time $R/c$, where $R$ is the size of the emitting
region, assumed spherical. 
As discussed in Ghisellini \& Tavecchio (2009), this assumption, suggested by the 
fast variability of blazars, allows { us} to neglect adiabatic losses, particle escape and 
the changed conditions in the emitting region: since the source is traveling 
and expanding, the magnetic field and the particle density do not dramatically change
in a time $R/c$.
}

The injected distribution, of total power $P^\prime_{\rm i}$ 
{ 
(primed quantities are calculated in the comoving frame of the source), 
is assumed to extend between $\gamma_{\rm min}=1$ and $\gamma_{\rm max}$
and to be a  broken power law  smoothly joining at $\gamma_{\rm b}$:
}
\begin{equation}
Q(\gamma)  \, = \, Q_0\, { (\gamma/\gamma_{\rm b})^{-s_1} \over 1+
(\gamma/\gamma_{\rm b})^{-s_1+s_2} }\, \quad [{\rm cm^{-3}\, s^{-1}}]
\label{qgamma}
\end{equation}
The normalization $Q_0$ is set through 
$P^\prime_{\rm i}=(4\pi /3) R^3 \int Q(\gamma)\gamma m_{\rm e}c^2 d\gamma$.
{
The emitting region is assumed to be located at a distance $R_{\rm diss}$
from the black hole. 
Its size is  $R=\psi R_{\rm diss}$, where $\psi$
is the semi--aperture angle of the jet, assumed conical.
We assume $\psi=0.1$ rad.
}
The model accounts for the accretion disk component, as well as for the
IR emission reprocessed by a dusty torus and the X--ray 
emission produced by a hot thermal corona placed above and below the accretion disk.
We have assumed that the accretion disk contributes significantly to the bluest 
fluxes observed by GROND, and this fixes both the disk luminosity $L_{\rm d}$
and the black hole mass $M$. 
We find a black hole mass $M=1.1\times 10^9 M_\odot$ and $L_{\rm d}=6.5\times 10^{45}$ erg s$^{-1}$.
{ This model} under--reproduces the red part of the  GROND SED,
which by itself may be fit with a simple power law.
This apparent excess may be caused by the { oversimplified} torus structure assumed by the model, 
or by some synchrotron emission produced by another component.
The ratio of the inverse Compton to synchrotron luminosity
(the so called Compton dominance) is rather large (factor $\sim$100), in agreement
with other powerful blazars. 
This suggests that the inverse Compton flux benefits from the presence of seed
photons produced not only by the synchrotron process (internal to the jet),
but also on photons produced externally to the jet, such as the 
broad line photons and the IR emission produced by the torus.
As {\it NuSTAR}  demonstrates, the X--ray spectrum 
is intrinsically very hard, and not because of absorption.
This indicates that the seed photons coming
from the broad line region and the torus are important as seeds for the formation
of the high energy bump, since the synchrotron
self--Compton \citep[SSC,][]{maraschi92} process would produce a softer
and less powerful luminosity (see the blue dotted line in Figure~\ref{sed}).
{
The large Compton dominance favors two specific locations
\citep[see][]{ghisellini09c,sikora09}:
i) the first is within the broad line region (``BLR" case), and 
ii) the second is
outside it, but within the torus (``torus" case).
These are the locations where the ratio between the radiation and magnetic energy
densities are as large as needed to explain the Compton dominance of the source.
However, the size of the emitting zone would be quite different in the two cases, 
and this corresponds to two different minimum variability time--scales,
which for our models are about one day for the ``BLR" case and five times
longer in the ``torus" case.
The two models produce very similar SED, with similar total jet power,
even if the bulk Lorentz factors, the injected power and the magnetic fields
are different.
Therefore the most promising way to distinguish is through variability
of the X and $\gamma$--ray fluxes, that are not contaminated by the
much steadier contributions of the disk and torus radiation.
}
%{ location} 
%of the emitting region is { closer to the black hole} than 
%{ the distance of} the BLR ($R_{\rm diss}<R_{\rm BLR}$).
%This is because, within $R_{\rm BLR}$, the radiation energy density 
%is larger than { that} produced by the torus (for the same bulk Lorentz factor).

{
If the radiation produced externally to the jet is important, as here, the beaming pattern of 
the synchrotron radiation is different from the beaming pattern of the inverse Compton process,
with the latter more enhanced in the forward direction \citep{dermer95}.
We take this effect into account.
Furthermore, we do not assume $\Gamma\sim \delta$, but treat both $\Gamma$ and the
viewing angle $\theta_{\rm v}$ as parameters of the model.
}

The obtained parameters, listed in Table~\ref{para}, are well within the range of parameters
found for other blazars of similar shape and Compton dominance
studied and interpreted with the same model
\citep{ghisellini10,ghisellini10b,ghisellini15}.
 Since the emitting region is rather compact 
{
($R=2.4\times 10^{16}$ cm in the ``BLR" case and 5 times that for the ``torus" case), 
} 
its radio emission is self--absorbed
(up to $\sim$400\,GHz) and cannot account for the observed radio flux, that must necessarily
come from much larger zones.
{
The total power $P_{\rm r}$ of the emitted bolometric luminosity 
is of the order \citep[see e.g.][]{ghisellini09c,ghisellini14b}
\begin{equation}
P_{\rm r} \, \sim { L_{\rm jet}^{\rm bol} \over \Gamma^2} \,  \sim\,  4 \times 10^{46}\quad  {\rm erg\, s^{-1}}
\end{equation}
}
This can be considered { to be} a lower limit { on} the total jet power.
This cannot be provided by the bulk motion of the relativistic emitting electrons nor by the
Poynting flux 
{ 
(see the corresponding values $P_{\rm e}$ and $P_{\rm B}$ in Table \ref{para},)
and requires the presence of an important proton component { that is} dynamically dominant.
The value reported in Table \ref{para}, assuming one cold proton per emitting electron,
is $P_{\rm p}\sim 4\times 10^{47}$ erg s$^{-1}$, a value much larger (factor 60) than $L_{\rm d}$.
Assuming a 10\% accretion efficiency, this would imply that $P_{\rm jet} \sim 6 \dot M c^2$.
One can lower $P_{\rm p}$ by assuming that there are some emitting e$^\pm$ pairs, but their number
cannot exceed $\sim$10 per proton. 
In this case $P_{\rm jet}\sim P_{\rm r}$, the
entire kinetic energy would be used to produce the radiation we see, and the jet would stop.
This limit the possible number of pairs to a few per proton 
\citep[see discussion in][]{sikora00,celotti08}.
}

% -------------------------------------------------------------------------------
\begin{figure*}[ht!]
\vskip -1 cm
\begin{center}
\includegraphics[scale=0.75]{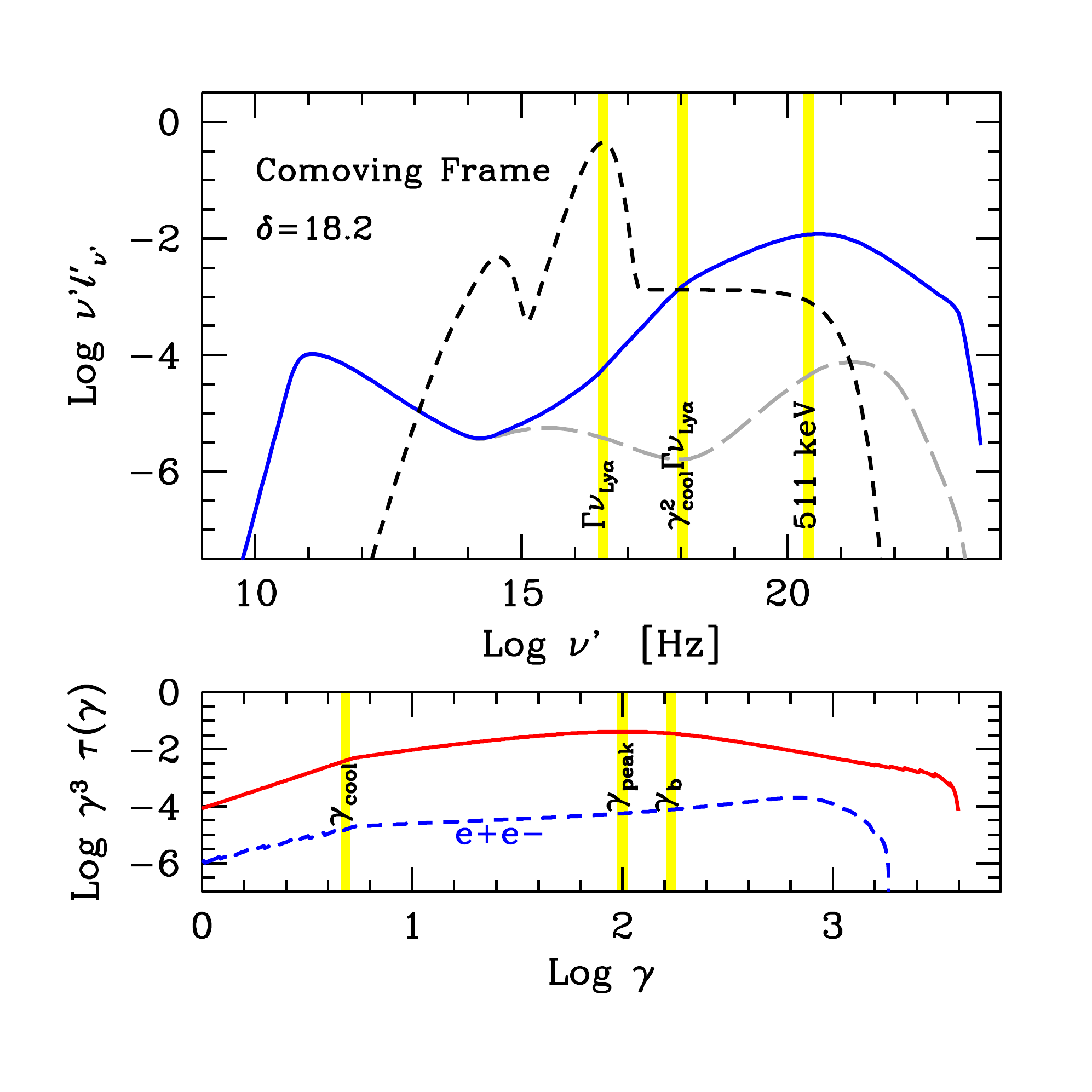}
\end{center}
\vskip -2 cm
\caption{
Top panel: SED of PMN~0641--0320 in the comoving frame (blue solid line). 
The black dashed line shows the spectrum of the IR torus, 
{ 
the BLR \cite[assumed to have a black--body shape, see][]{tavecchio08}
}
and the X--ray corona, while the gray long
dashed line shows the synchrotron self Compton component. 
The units are $\nu^\prime\ell^\prime(\nu^\prime)$, where $\ell^\prime(\nu^\prime)$ 
is the monochromatic compactness defined as 
$\ell^\prime(\nu^\prime)\equiv \sigma_{\rm T}L^\prime(\nu^\prime)/[R m_{\rm e}c^3]$.
In this frame the observer sees an enhanced BLR and torus component.
The inverse Compton scattering, in this frame,
can use the entire amount of seed photons only for $\nu^\prime > \Gamma \nu_{\rm Ly\alpha}$,
which becomes $\nu=\delta\nu^\prime/(1+z)$ in the observed frame.
The bottom panel shows the particle distribution in the form 
$\gamma^3\tau(\gamma)\equiv \gamma^3 \sigma_{\rm T} R N(\gamma)$, 
resulting from the solution of the continuity equation,
that accounts for injection, radiative cooling, and pair production.
Below $\gamma_{\rm cool}$ the electrons do not cool in the light crossing time $R/c$,
and the particle distribution retains the same slope $s_1$ of the injection function.
Above the break energy $\gamma_{\rm b}$ the slope of $\tau(\gamma)$ is $s_2+1$.
The $\gamma^3\tau(\gamma)$ representation allows to easily find $\gamma_{\rm peak}$,
i.e. is the energy producing the two peaks of the SED.
The dashed line shows the (in this case modest) contribution of the electron--positron pairs
produced { within} the emitting region.
\label{sedcom}}
\end{figure*}
% -------------------------------------------------------------------------------

\subsection{The X--ray hardness}

As mentioned above, the extremely hard slope of the X-ray spectrum,
that cannot be due to absorption, strongly suggests that the inverse Compton
process uses external photons as seeds.
This is due to two reasons:

i) In the top panel of Figure~\ref{sedcom}, we show the SED as seen by an observer comoving 
with the emitting blob.
We use the $\nu^\prime \ell^\prime(\nu^\prime)$ versus $\nu^\prime$ representation,
where $\ell^\prime(\nu^\prime)$ is the monochromatic compactness, defined as 
$\ell^\prime(\nu^\prime)\equiv \sigma_{\rm T}L^\prime(\nu^\prime)/[R m_{\rm e}c^3]$ \citep{cavaliere1980}.
Since $\nu^\prime \ell^\prime(\nu^\prime)$ is a measure of the optical depth for the pair--production process
{ 
(becoming important for $\nu^\prime \ell^\prime(\nu^\prime)>1$),
}
the top panel of Figure~\ref{sedcom} shows that pair production is marginal.
{
This is confirmed by the relatively small amount of electron--positron pairs
produced by the $\gamma$--$\gamma \to$e$^\pm$ process, shown
in the bottom panel of Figure~\ref{sedcom}, together with the original primary electrons.
}

{
In the comoving frame of the emitting blob,
}
the photons produced by the disk, the BLR and the torus
are seen Doppler shifted and aberrated by different amounts, depending on
the angle between the photon direction and the velocity vector.
{ 
For our values of $R_{\rm diss}$,  
most of the disk radiation, produced by the inner part of the disk,
would be seen redshifted in the comoving frame of the blob, and does not 
contribute much to the seed photons for the inverse Compton scattering process.
Much more important are the seeds produced by the BLR and the torus.
As long as the blob is inside the BLR (torus), the emission of the BLR (torus) is
seen beamed, with a corresponding energy density enhanced by a factor $f \Gamma^2$
with respect to an observer stationary with respect to the black hole.
{
The $f$ parameter is of order unity, as long as $R_{\rm diss}< R_{\rm BLR}$ (for the BLR case)
or $R_{\rm diss}< R_{\rm torus}$ (torus case) and 
its exact value depends on the geometry of the BLR (spherical or flattened) and the torus.
} 

The top panel of 
}
Figure \ref{sedcom} illustrates the ``BLR" case as shown in Figure \ref{sed}, 
{
but the plotted SED is as observed in the comoving frame of the blob.
}
Most of the seed photons are provided by the BLR, and especially the 
hydrogen Ly$\alpha$. 
The frequency of these photons is seen (in the frame comoving with the blob) 
at $\nu^\prime_{\rm seed}\sim \Gamma \nu_{\rm Ly\alpha}\sim 3\times 10^{16}$ Hz.
Below $\nu^\prime_{\rm seed}$ the inverse Compton process can scatter seed photons of lower
frequencies, {\it that are fewer in number}.
Scattering with relatively cold electrons of $\gamma\sim 1$, photons at  $\nu^\prime_{\rm seed}$ 
will remain at the same frequency in the comoving frame, but will be observed 
at $\delta \nu^\prime_{\rm seed}$, that in our case is of the order of $\sim$2.5 keV.
As a consequence, the { resulting} inverse Compton spectrum is predicted to be hard below this frequency,
because of the relative paucity of seed photons below $\nu^\prime_{\rm seed}$.
 
On the other hand the X--ray spectrum of the source continues to be very hard up to $\sim 70$ keV.
Therefore, this explanation is not sufficient to account for the hardness across the entire observed
X--ray energy range. We thus suggest an additional reason:

ii) The inverse Compton process efficiently cools the electrons.
{
Electrons above $\gamma_{\rm cool}\sim 5$ radiatively cool in one light crossing time
(i.e. they halve their energy). 
Electrons below this energy radiatively cool in a longer time,
and will be affected by adiabatic cooling
(important after a doubling time of the source).
Assuming that the injection stops after one light crossing time,
and calculating the SED at this time,
we find that the particle distribution
% Electrons with energies below $\gamma_{\rm cool}$ do not cool, and the corresponding distribution 
$N(\gamma)$, below $\gamma_{\rm cool}$,
}
retains the injection slope (which is hard in our case: $s_1=0.5$ 
[namely $N(\gamma)\propto \gamma^{-0.5}$ below $\gamma_{\rm cool}$]).
This is illustrated in the bottom panel of Figure \ref{sedcom}, 
showing $\gamma^3 \tau(\gamma)$ as a function of $\gamma$ 
[where $\tau(\gamma) \equiv \sigma_{\rm T}RN(\gamma)$].
The $\gamma^3$ factor allows { us} to immediately see what electron 
energies contribute the most at the two peaks of the SED.
The very hard electron distribution in the range $1<\gamma<\gamma_{\rm cool}\sim$5 corresponds 
to a very hard spectrum, up to $\delta\gamma^2_{\rm cool}\nu^\prime_{\rm seed}\sim$ 75 keV.

These two { factors} act together to harden the slope of the X--ray spectrum, making it harder
than $F(\nu)\propto \nu^{-0.5}$ (equivalent to $dN/dE\propto E^{-1.5}$), that would be typical for fast cooling electrons below $\gamma_{\rm b}$
scattering a fixed amount of { soft seed} photons.
%The peak of the Compton part of the SED is constrained to lie in the 
% MeV band by the hard X--ray and the very soft $\gamma$--ray slopes. 
{\it NuSTAR} { also fixes} the X--ray slope up to $\sim$70 keV, and  the very soft
{\it Fermi}--LAT spectrum constrains the peak of the Compton component
to lie in the MeV band. 
In turn, this constrains both $\gamma_{\rm b}$ and $\gamma_{cool}$ to
be smaller than $\sim$45.
If the emitting region were at much larger distances from the black hole, 
with no external photons, we would have the problem { of explaining} 
the large Compton dominance, and also how electrons with $\gamma>45$ 
cool efficiently.

{
Very similar considerations can be done for the torus case.
In this case $\gamma_{\rm cool}\sim 109$ is larger, but the seed photon frequency
(the peak of the IR torus emission) is smaller (we assume a temperature of 370 K)
leading to approximately the same inverse Compton frequency peak (see Fig. \ref{sed}).
}

%%%%%%%%%%%%%%%%%%%%%%%%%%%%%%%%%%%%%%%%%%%%%%%%%%%%%%%%%%%%%%%%%%%%%%%%%%
%
% Conclusions
%
%%%%%%%%%%%%%%%%%%%%%%%%%%%%%%%%%%%%%%%%%%%%%%%%%%%%%%%%%%%%%%%%%%%%%%%%%%

\section{Conclusions}
`MeV blazars' are the most powerful type of blazars and among the most luminous persistent
sources in the Universe. Their large jet power, accretion luminosity, and black hole mass set them
apart from the rest of jetted AGN. Despite their high luminosity, only
a handful of bona-fide `MeV blazars' were known { until} recently \citep{bloom96,collmar06,sambruna06,ajello09} 
because of the lack of an MeV telescope surveying the entire sky.
However, MeV blazars are characterized by an extremely hard (power--law index $<$1.5) X--ray continuum
and the launch of {\em NuSTAR} has  uncovered a few new members of the MeV blazar
family \citep{sbarrato13,tagliaferri2015}.

In this paper we report on  ToO observations performed by {\em NuSTAR}, {\it Swift}
and GROND of a flaring source, Fermi J0641$-$0317,  
detected by {\it Fermi}--LAT in the direction of the anti--center of our Galaxy. 
These observations showed that the counterpart of  Fermi J0641$-$0317
is PMN~J0641$-$0320 a very bright (8.6\,GHz flux of 0.83\,Jy) radio source,
which, our Keck observation places at a redshift of $z=1.196$.

The overall SED of PMN~J0641$-$0320, built with contemporaneous and
 semi--simultaneous observations, unveils several important characteristics.
First,  PMN~J0641$-$0320 displays, while flaring, the SED of a powerful
blazar with a peak luminosity of $L\geq 10^{48}$\,erg s$^{-1}$, a high-energy
peak located in the MeV band and a Compton dominance of a factor $\sim$100.
Second, the large Compton dominance suggests that most of the high-energy emission
is produced via inverse Compton scattering of the accelerated electrons off 
an external photon field, very likely the BLR and/or the infrared torus.
Our SED modeling suggests a black hole mass of $\sim10^9$\,M$_{\sun}$.

The X--ray continuum, which {\em NuSTAR} detects and characterizes up to 70\,keV (150\,keV in the source frame),
is extremely hard and can be characterized by a power law with a photon index of $\Gamma_X\approx$1. 
This makes  PMN~J0641$-$0320 one of the hardest X--ray emitting
blazars and one of the hardest {\it NuSTAR} sources.
The extreme X--ray hardness is interpreted, in the framework of the external inverse Compton scenario, 
as produced by a hard electron distribution, which below $\gamma_{\rm cool}\approx$5
is not cooled, and retains the shape of the injected spectrum, $N(\gamma)\propto\gamma^{-0.5}$,
and thus causes the very hard X--ray spectrum. 

The jet radiative power ($P_{\rm r}$ in Table~\ref{para}), which is a lower limit to the true jet power, 
is larger than the disk luminosity ($L_{\rm d}$ in Table~\ref{para}), 
which suggests that the jet is not only powered via accretion, but taps into the rotational 
energy of the spinning black hole as found for other powerful blazars \citep{ghisellini14b,tagliaferri2015}.

The hard X--ray continuum, the SED peak location, the large Compton
dominance and the high luminosity 
identify PMN~J0641$-$0320 as a new member of the  MeV blazar family. 
MeV blazars may substantially contribute to the MeV background \citep{ajello09} and can be used to constrain 
the mass density of heavy black holes \citep{ghisellini10,sbarrato14}.
The analysis of $\sim$6\,years of {\it Fermi}--LAT data shows that PMN~J0641$-$0320
underwent two rather long flaring episodes. 
Indeed, it is not unusual for MeV blazars to flare for weeks at a time in $\gamma$ rays. 
This together with the increased
sensitivity, due to Pass~8, of {\it Fermi}--LAT at $<$100 MeV might allow us to uncover, 
in { combination} with {\em NuSTAR} observations, new powerful blazars.

%%%%%%%%%%%%%%%%%%%%%%%%%%%%%%%%%%%%%%%%%%%%%%%%%%%%%%%%%%%%%%%%%%%%%%%%%%
%%%%%%%%%%%%%%%%%%%%%%%%%%%%%%%%%%%%%%%%%%%%%%%%%%%%%%%%%%%%%%%%%%%%%%%%%%
\clearpage
\acknowledgments
MA acknowledges generous support from  NASA grant NNH09ZDA001N.
MB acknowledges support from the International Fulbright Science and
Technology Award
and from NASA Headquarters under the NASA Earth and Space Science Fellowship Program, grant NNX14AQ07H

The \textit{Fermi} LAT Collaboration acknowledges generous ongoing support
from a number of agencies and institutes that have supported both the
development and the operation of the LAT as well as scientific data analysis.
These include the National Aeronautics and Space Administration and the
Department of Energy in the United States, the Commissariat \`a l'Energie Atomique
and the Centre National de la Recherche Scientifique / Institut National de Physique Nucl\'eaire 
et de Physique des Particules in France, the Agenzia 
Spaziale Italiana and the Istituto Nazionale di Fisica Nucleare in Italy, 
the Ministry of Education, Culture, Sports, Science and Technology (MEXT), 
High Energy Accelerator Research Organization (KEK) and Japan Aerospace 
Exploration Agency (JAXA) in Japan, and the K.~A.~Wallenberg Foundation, 
the Swedish Research Council and the Swedish National Space Board in Sweden.
Additional support for science analysis during the operations phase 
is gratefully acknowledged from the Istituto Nazionale di Astrofisica in 
Italy and the Centre National d'\'Etudes Spatiales in France.

This {\it NuSTAR} work was supported under NASA Contract No.~NNG08FD60C, and made use of data from the 
{\em NuSTAR} mission, a project led by the California Institute of Technology, managed by the 
Jet Propulsion Laboratory, and funded by the National Aeronautics and Space  Administration. 
We thank the {\em NuSTAR} Operations, Software and  Calibration teams for support with the 
execution and analysis of these observations. This research has made use of the {\em NuSTAR} 
Data Analysis Software (NuSTARDAS) jointly developed by the ASI Science Data Center (ASDC, Italy) 
and the California Institute of Technology (USA).

Part of this work is based on archival data, software or on--line services 
provided by the ASI Data Center (ASDC).
This research has made use of the XRT Data Analysis Software (XRTDAS).
Part of the funding for GROND (both hardware and personnel) 
was generously granted by the Leibniz-Prize to G. Hasinger (DFG grant HA 1850/28-1).

{\it Facilities:} \facility{Fermi/LAT}, \facility{{\it NuSTAR}},
\facility{{\it Swift}}, \facility{GROND}

%%%%%%%%%%%%%%%%%%%%%%%%%%%%%%%%%%%%%%%%%%%%%%%%%% biblio
\bibliographystyle{apj}
\bibliography{/Users/majello/Work/Papers/BiblioLib/biblio.bib}

\end{document}